\documentclass[unsortedaddress,preprint,aps,show pacs,showkeys]{revtex4}
\usepackage{graphicx}
\usepackage{dcolumn}
\usepackage{amsmath}
\usepackage{amssymb}
\usepackage{amsfonts}
\usepackage{bm}
\usepackage{color}
\newcommand{\be}{\begin{equation}}
\newcommand{\ee}{\end{equation}}
\newcommand{\ba}{\begin{eqnarray}}
\newcommand{\ea}{\end{eqnarray}}
\begin{document}
\title{Twofold reentrant melting in a double-Gaussian fluid}
\author{Santi Prestipino\footnote{Corresponding author. Email: {\tt sprestipino@unime.it}}, Cristina Speranza\footnote{Email: {\tt csperanza@unime.it}}, Gianpietro Malescio\footnote{Email: {\tt malescio@unime.it}}, and Paolo V. Giaquinta\footnote{Email: {\tt paolo.giaquinta@unime.it}}}
\affiliation{Universit\`a degli Studi di Messina, Dipartimento di Fisica e di Scienze della Terra, Contrada Papardo, I-98166 Messina, Italy}
\date{\today}
\begin{abstract}
Isotropic pair potentials that are bounded at the origin have been proposed from time to time as models of the effective interaction between macromolecules of interest in the chemical physics of soft matter. We present a thorough study of the phase behavior of point particles interacting through a potential which combines a bounded short-range repulsion with a much weaker attraction at moderate distances, both of Gaussian shape. Notwithstanding the fact that the attraction acts as a small perturbation of the Gaussian-core model potential, the phase diagram of the double-Gaussian model (DGM) is far richer, showing two fluid phases and four distinct solid phases in the case that we have studied. Using free-energy calculations, the various regions of confluence of three distinct phases in the DGM system have all been characterized in detail. Moreover, two distinct lines of reentrant melting are found, and for each of them a rationale is provided in terms of the elastic properties of the solid phases.
\end{abstract}
\maketitle

\section{Introduction}

Interaction potentials accounting for effective forces in complex fluids can be quite different from those typical of atomic fluids~\cite{Likos}. In particular, while in elementary fluids the short-range interaction is always infinitely repulsive due to excluded-volume effects, in complex fluids the effective interactions obtained by averaging over microscopic/internal degrees of freedom may result in a bounded repulsion, thus allowing full particle interpenetration. For example, polymers in a good solvent form highly penetrable coils and the repulsion between their centers of mass is finite at all distances, decaying rapidly beyond the radius of gyration of the coils. For self-avoiding polymers, the effective pair potential is reasonably well represented by a Gaussian whose width is of the order of the radius of gyration whereas its value at zero separation is roughly $2k_BT$~\cite{Louis}.

A bounded repulsion may be combined with an attractive tail. For instance, in polymer solutions depletion (e.g. Asakura-Oosawa) forces give rise to a weak attraction between two polymers~\cite{Likos}. However, a pair potential with a bounded repulsion and an attractive component may be thermodynamically unstable. The issue of thermodynamic stability in the context of the penetrable-sphere potential supplemented by a square-well attraction has been recently discussed by Giacometti and coworkers~\cite{Fantoni1,Fantoni2}. As first pointed out by Fisher and Ruelle~\cite{Fisher}, if the potential is not sufficiently repulsive to discourage particles overlap, the system collapses into a small volume and no thermodynamic limit exists. Fisher and Ruelle derived sufficient conditions for a bounded pair repulsion with an attractive tail to lead to thermodynamic equilibrium. 

In the present paper we investigate the phase behavior of a model fluid with a bounded interparticle repulsion and a longer-ranged attraction. In particular, both the repulsion and the attraction are given a Gaussian shape. Our purpose is to investigate the effect of a weak attraction on the melting behavior associated with the Gaussian repulsion. The strength of the attraction is chosen small enough so that the double-Gaussian fluid is everywhere thermodynamically stable. Moreover, the center of the attractive well is displaced from the repulsive core in order that the strength of the repulsion be roughly unaffected by the attraction. This makes it possible to compare the phase behavior of the system investigated here with that of the popular Gaussian-core model (GCM) fluid~\cite{Stillinger,Lang,Prestipino1,gcm2d,gcm1d}.

The outline of the paper is the following. After introducing the double-Gaussian model and briefly describing the simulation method in Section II, our results are presented in Section III. Thanks to these calculations, the characteristics of the phase diagram of the system are uncovered to their finest details. Some concluding remarks are given in Section IV.

\section{Model and method}
\setcounter{equation}{0}
\renewcommand{\theequation}{2.\arabic{equation}}

We are going to explore how the phase diagram of the three-dimensional GCM fluid is being modified when a shallow attractive well is added, at medium-range distances, to the repulsive Gaussian core. Clearly, a prominent effect of the attraction will be to provide the original GCM system with another (liquid, as distinct from vapor) fluid phase, although we also expect important changes in the solid sector of the phase diagram, at least for pressures low enough that the average distance between two neighboring particles in the system is much larger than the core size.

To be concrete, let the following pair potential be considered:
\be
u(r)=\epsilon\exp\{-r^2/\sigma^2\}-\epsilon_2\exp\{-(r-\xi)^2/\sigma_2^2\}\,,
\label{2-1}
\ee
where $\epsilon$ and $\sigma$ are arbitrary energy and length units, respectively. The potential (\ref{2-1}) defines what will be called the general double-Gaussian model (DGM). As far as we know, no thorough study of the phase behavior of the DGM potential has hitherto been performed, nor even for $\xi=0$. Aiming to perturb the GCM phase diagram only slightly, we shall take, quite arbitrarily, $\epsilon_2=0.005\,\epsilon,\sigma_2=\sigma$, and $\xi=3\sigma$ (see Fig.\,1). With this choice, the inner particle core is almost unchanged with respect to the GCM with the same $\epsilon$ and $\sigma$, thus suggesting very similar GCM and DGM high-pressure behaviors; moreover, considering the closeness of $\epsilon_2/k_B$ to the maximum GCM melting temperature $T_{\rm max}$, we expect the critical temperature $T_c$ of the DGM fluid to be of the order of $T_{\rm max}$ too.

As anticipated in the Introduction, a problem of stability may arise for a system of particles interacting through a repulsive potential that is finite at the origin when it is augmented with an attractive tail. If the added attraction is too strong, a thermodynamic catastrophe occurs in that, in the infinite-size limit, all system particles eventually concentrate in a {\em finite} region of space; should this happen, no clearly-defined thermodynamics would be possible for the system. In order to figure out whether this be the case or not for the DGM potential hereby considered, we resorted to a pair of criteria originally put forward by Ruelle (see Propositions 3.2.4 and 3.2.7 of Ref.\,\cite{Ruelle}), which have been recently revived by Heyes and Rickayzen (Theorems 1 and 2 of Ref.~\cite{Heyes}). Denoting $\widetilde{u}(k)$ the Fourier transform of the potential, a sufficient condition for instability is $\widetilde{u}(0)<0$. Conversely, if we are able to prove that $\widetilde{u}(k)\ge 0$ for all $k$ then the system is stable. For the interaction potential in Eq.\,(\ref{2-1}), the Fourier transform can be formally evaluated in terms of the error function of complex argument. For $\sigma_2=\sigma$ and $\xi=3\sigma$, we find $\widetilde{u}(0)<0$ for $\epsilon_2>0.026315\,\epsilon$. Below exactly the same threshold $\widetilde{u}(k)$ is positive definite. Hence, we were able to decide on the stability of the system for all values of $\epsilon_2$: for $\sigma_2=\sigma$ and $\xi=3\sigma$, the DGM system is stable for $\epsilon_2<0.026315\,\epsilon$; above this value, it is actually unstable and thus $u(r)$ cannot serve as an interaction potential of stable matter.

Once the stability of the DGM system has been established, the next step is to find the relevant crystalline structures at low temperatures, so as to provide the necessary input to the simulation of the system in the solid sector of the phase diagram. This was accomplished through exact $T=0$ total-energy calculations for a set of candidate structures comprising Bravais and non-Bravais lattices with at most one internal parameter (see the complete list of such structures in Ref.\,\cite{Prestipino2}). This list also included clustered fcc and bcc crystals with two particles per lattice site~\cite{gcm1d}.

Table 1 reports the $T=0$ phase boundaries for those solids whose chemical potential turned out to be lowest over a range of pressures (any other crystalline phase is much far above in chemical potential to be considered relevant for non-zero temperatures). We see that there are two distinct $P$ intervals where the bcc phase is thermodynamically stable at $T=0$. In particular, a bcc solid is expected to occur at very low pressures. Although we can easily envisage the existence of a subtle competition between this dilute solid and the fluid phases, this question can only be settled by accurate free-energy calculations.

%
%
\begin{table}
\caption{Zero-temperature phases of the DGM system under study. For each pressure range in column 1, the thermodynamically stable phase is indicated (column 3) together with its values of the number density $\rho$ (column 2).}
\begin{tabular*}{\columnwidth}[c]{@{\extracolsep{\fill}}|c|c|c|}
\hline
$P$ range ($\epsilon\,\sigma^{-3}$) & $\rho$ range ($\sigma^{-3}$) & stable phase \\
\hline\hline
0-0.00013 & 0.0677-0.0710 & bcc \\
\hline
0.00014-0.01162 & 0.0781-0.1361 & fcc \\
\hline
0.01163-0.03386 & 0.1364-0.1752 & hcp \\
\hline
$\ge 0.03387$ & $\ge 0.1761$ & bcc \\
\hline
\end{tabular*}
\end{table}

The phase diagram of the DGM model was carefully investigated by Monte Carlo (MC) simulation in the $NPT$ (isothermal-isobaric) ensemble (Section III). In order to trace the liquid-vapor binodal line, we made use of Gibbs-ensemble simulations~\cite{Panagiotopoulos}. Systems of about 1000 particles (with periodic repetition of the simulation box) are perfectly suitable for investigating bulk properties and for determining phase boundaries; for such sizes, there is no advantage in employing cell linked lists in the simulation. Typically, as many as $10^5$ sweeps/cycles were generated at equilibrium for every $(T,P)$ point along a simulation path, which proved sufficient to obtain accurate statistical averages for the volume and the energy per particle. Much longer runs of $5\times 10^5$ sweeps each were performed for the calculation of the chemical potential in the fluid phase by Widom's particle-insertion method~\cite{Widom}. The location of the melting transition was determined through thermodynamic integration along isobaric and isothermal paths (see e.g. Ref.~\cite{Saija}), combined with ``exact'' free-energy calculations for some selected states. While in the fluid sector of the phase diagram the reference state was a dilute gas, in the solid region we chose a low-temperature crystal (a different one for each solid phase) as the starting point of a MC trajectory. In any such state, the excess Helmholtz free energy per particle was computed by the Einstein-crystal method~\cite{Frenkel,Polson}.

\section{Results}
\setcounter{equation}{0}
\renewcommand{\theequation}{3.\arabic{equation}}

We display in Fig.\,2 the overall $P$-$T$ phase diagram of the DGM system as derived from our numerical free-energy calculations. To improve the visibility of the low-pressure region, we have unevenly stretched the pressure scale by reporting the fourth root of $P$ rather than $P$ itself on the horizontal axis. The same phase diagram but in the $\rho$-$T$ plane is shown in Fig.\,3, so as to highlight the coexistence regions. Finally, in Fig.\,4 a zoom is made on the extremely-low-density region, which is where the main novelties over the GCM case are concentrated. In the appendix, the ``exact'' transition lines are compared with those derived by a number of (necessarily approximate) theoretical approaches.

By a first glance at Fig.\,2, one immediately realizes how complete and congruous is the thermodynamic picture emerging from our simulations. Except for the vapor-bcc coexistence locus, which is confined to such an extremely small neighborhood of $P=0$ that it could not be analyzed in detail, all the other transition lines in Fig.\,2 are smooth enough and well characterized to allow us to resolve the most minute details of the DGM phase diagram with precision. On the low-pressure/high-temperature side of Fig.\,2 we see the liquid-vapor coexistence line, which was traced by reporting the common value of the pressure in the two simulation boxes at the end of the Gibbs-ensemble runs. By the usual extrapolation procedure~\cite{Frenkel}, we locate the critical point at $\rho_c=0.02295$, $T_c=0.00767$, and $P_c=0.000046$ (from now on, all quantities will be given in reduced units). The liquid-vapor coexistence region is shown in Fig.\,3; interestingly enough, it fairly coincides with the region within which the hypernetted-chain (HNC) equation~\cite{Hansen} could not be solved.

Another expected feature of the DGM phase diagram is the reentrant melting of the high-pressure bcc phase, which occurs similarly as in the GCM: upon increasing $P$, the nearest-neighbor (NN) distance in the bcc solid becomes more and more blurred, due to an increasingly easier penetration of the first-shell particles into the finite repulsive core, until the coordination shells all crumble together at the melting point. Compared with the GCM, the maximum $T_{\rm max}$ of the DGM melting temperature is slightly larger whereas the pressure $P_{\rm max}$ in the same point is a little smaller.

The range of stability of the hcp crystal narrows upon heating, until the hcp phase ceases to exist as a stable phase at a temperature slightly smaller than 0.0045. Beyond this point, a fcc-bcc coexistence line starts, which seems the continuation of the hcp-bcc transition line towards higher temperatures; the fcc-bcc locus eventually terminates with roughly zero slope at the confluence with the melting line, which occurs slightly below $T=0.0050$ (this feature is vaguely reminiscent of the fcc-bcc coexistence line in the GCM, where however the point of maximum temperature falls within the solid sector of the phase diagram).

Let us now focus on the low-pressure region of the phase diagram, where according to our calculations a number of unconventional features occur for the DGM. To begin with, we have the indisputable evidence of a tiny region of bcc stability at low pressure. In this dilute crystal the average NN distance is close below the ``center'' of the attractive well, $\xi$. Moreover, the slope of the melting line is negative. On the low-density side the bcc-liquid coexistence locus meets the liquid-vapour line at a triple point; on the high-density side it merges with the fcc-liquid line at another triple point, which also represents the lowest-temperature state where the liquid exists as a stable phase. The bcc density along the sublimation line was estimated through $P=0$ simulations of the crystal (the actual coexistence pressures are indeed smaller than $10^{-6}$).

The DGM phase diagram of Fig.\,3 is also reproduced in part in Fig.\,4, showing a magnification of the low-density/low-temperature region. The most curious feature emerging from this picture has to do with the nature of the bcc-liquid equilibrium, which unexpectedly sees the bcc phase on the low-density (rather than on the high-density) side. Upon compression, and before the fcc crystal becomes relevant (which only occurs beyond a certain pressure), the NN distance in the bcc crystal is pushed increasingly farther away from $\xi$, with the effect of destabilizing the bcc phase with respect to the liquid phase. The net result will be a second reentrant-melting line (besides the one existing at high pressure), ending at a bcc-liquid-fcc triple point. Another explanation of this anomaly underscores the effect of pressure on the ``degree of rigidity'' of the dilute bcc phase, which in the pressure range under focus becomes progressively smaller with increasing compression (see Fig.\,8 in the appendix).

One aim of the present study was to ascertain the nature of the liquid phase in presence of a bounded repulsion between the particles. For example, in Fig.\,5 we plot the radial distribution function (RDF) of the DGM system for a number of temperatures, for $P=0.002$ (i.e., a typical value for the liquid). For comparison, also the RDF of the GCM fluid is plotted along the same isobar. There is clearly more structure in the DGM liquid than in the GCM fluid of same $T$ and $P$, owing to the stabilizing effect of the attraction which enhances the local density and overall improves the spatial definition of the coordination shells around a particle. At the chosen pressure of 0.002, there would anyway be little influence of the inner potential core on the RDF, other than providing a barrier against system collapse; in other words, particles would be blind to the exact shape of the inner core, i.e., to whether it is soft or hard, since the average separation between neighboring particles is close to $\xi$.

Finally, we explore how the phase diagram of the double-Gaussian fluid changes when $\epsilon_2$ and $\xi$ are slightly modified from the values hereby considered (for simplicity, we only admitted the fcc, bcc, and hcp crystals as candidate solids). Since there is no space for a detailed study here, we shall make use of the HNC equation in association with the melting criterion of Ref.\,\cite{Prestipino3}, which is shown in the appendix to predict very carefully the topology of the phase diagram for the case $\epsilon_2=0.005$ and $\xi=3$. Looking at the phase portraits of Figs.\,6 and 7, we soon realize that (i) a deeper attractive well will increase the size of the liquid-vapor coexistence region, with repercussions also on the sequence of stable solids with pressure (for example, for $\epsilon_2=0.015$ and 0.020 the bcc crystal is the only zero-temperature stable phase); (ii) as $\xi$ is increased, the nature of the anomalous melting becomes more complicated, with significant modifications in the shape of the melting line (for $\xi=4$, the fcc crystal is stable at $T=0$ in a range of densities between roughly 0.16 and 0.21; this is apparently confirmed also for non-zero temperatures). Focusing on Fig.\,6, we see that, as the strength of the Gaussian attraction becomes larger and larger, the liquid-vapor line extends to increasingly higher pressures and temperatures until it eventually passes over the point of maximum $T_m$ (this would occur for $\epsilon_2\approx 0.020\epsilon$ --- remember: the values of $T_m$ in Fig.\,6 would be overestimated by roughly $100\%$). If a suitably small hard core is added to this potential, a bcc-fcc phase transition will be induced at higher pressure, with likely little influence on the low-pressure part of the phase diagram. The overall phase portrait would now be reminiscent of that of water, with the bcc-fcc coexistence line mimicking the locus of points separating ice-I from ice-III in water. Also the waterlike anomalies would be located at the right place, i.e., above the unique reentrant-melting line. The actual effectiveness of this approach to generating an isotropic potential with a phase behavior reminiscent of water is currently under investigation.

\section{Conclusions}

In this paper, we carefully analyzed the phase behavior of a system of particles interacting through an isotropic two-body potential combining a Gaussian repulsion with a much weaker Gaussian attraction. A weak attraction at medium distances would be a generic trait of the pair interaction between fully interpenetrating polymer chains in a solution, whence the importance of understanding the effects of small attractive forces on the thermodynamic behavior of softly repulsive particles. We found that even a modest attraction is able to change the phase diagram of the GCM in important ways. First of all, a different solid polymorphism shows up in the DGM system, with novel hcp and bcc phases. Of these, the (low-density) bcc phase shows a form of reentrant melting which has a completely different origin from the anomalous melting of the other (high-density) bcc phase. While the latter GCM-type melting is ultimately determined by the boundedness of the repulsive core, it is the shallow potential well that plays a leading role in the weakening of crystal coherence upon compression observed in the low-density bcc phase. In simpler terms, on approaching the bcc-fcc transition pressure at $T=0$, the rigidity of the bcc crystal progressively reduces as a result of the increasing detuning of the nearest-neighbor distance from the center of the attractive well.

We have finally noted that a promising approach to building up an isotropic water-type fluid would be to first increase the strength of the Gaussian attraction and then supplement it with a suitably chosen hard core. The feasibility of this scenario will be the subject of a future study.

\section*{Acknowledgements}
We gratefully acknowledge fruitful discussions with Ezio Bruno and Dino Costa.

\appendix
\section{Heuristic approaches}
\setcounter{equation}{0}
\renewcommand{\theequation}{A.\arabic{equation}}

In this appendix, we outline a number of theoretical approaches which aim at gaining some information on the overall structure of the DGM phase diagram, prior to carrying out the simulation.

Some useful hints on the phase-diagram topology come from a calculation that needs only a few minutes to be performed on a desktop computer. It makes use of the semi-empirical melting criterion introduced in Refs.\,\cite{Prestipino3,Prestipino4}, which combines the Lindemann melting rule with a description of the solid phase as an elastic continuum. This method proved to be effective in many cases, at least in predicting the exact topology of the melting line; on the other hand, for all the potentials investigated so far the melting temperature $T_m$ was systematically overestimated by roughly a factor of two. In practice, $T_m$ is first estimated as a function of pressure or density for a number of crystalline phases; then, the alleged melting-temperature locus is the upper envelope of all the individual melting lines drawn. Fig.\,8 shows the melting lines of the fcc, bcc, and hcp phases as computed by this method for the GCM fluid (left) and for the present DGM system (right). Aside from the erroneous $T_m$ scale, the exact GCM phase diagram~\cite{Prestipino1} is clearly apparent in the left panel of Fig.\,8. From the same plot we draw the prediction of a fcc phase becoming, upon compression, eventually superseded by a bcc crystal, which turns out to be correct. Upon switching on a weak attraction at moderately large distances, the GCM phase diagram gets modified only at the lowest densities (say, below 0.1 in reduced units). A second bcc basin appears at low pressures, which is partially hidden under the liquid-vapor coexistence region as approximated through the region where the HNC equation has no solution; within the region of stability of this bcc phase, the melting line would be a descending function of pressure. At $\rho=0.0742\,\sigma^{-3}$, the melting temperature falls to zero since right at this density the bcc crystal loses rigidity against tetragonal shear\,\cite{Prestipino3}. In the same region of densities, the DGM system would exist as a liquid down to very low temperatures, bounded by bcc states on the low-density side and by fcc states on the high-density side. Indeed, two further triple points (i.e., vapor-bcc-liquid and bcc-liquid-fcc) beyond the one also present in the Gaussian fluid do appear in the DGM phase diagram (Sect. III), and the melting line joining them is a further reentrant-melting line.

More reliable estimates of DGM melting points are obtained by the so-called heat-until-it-melts (HUIM) method~\cite{Saija2}. Following the HUIM approach, several heating trajectories of the solid are generated by MC simulation at selected pressures or densities, and eventually terminated at the point where the sample is observed to melt spontaneously (an event signalled by a distinct jump in both system energy and density). The accuracy of the melting temperature estimated in this way is usually poor (a solid can be overheated a lot, even by a $20\%$ of $T_m$, and to varying degrees as a function of pressure), but nonetheless the outcome  gives a clear indication of the trends of $T_m$ with pressure. From these calculations (see the data plotted in Figs.\,2 and 4) we learn that (i) reentrant melting in the DGM system occurs at high pressure in almost the same terms as in the GCM fluid; (ii) the melting curve of the (possibly metastable) low-pressure bcc phase would be almost flat.

Yet another approach is the variational theory (VT), which is an adaptation to the DGM of the same approach followed for the GCM by the authors of Ref.\,\cite{Lang} (see also Ref.\,\cite{Likos}). The idea is to combine the HNC virial route for the fluid phase with an approximation for the Helmholtz free energy $F$ of the crystal based on the use of the Gibbs-Bogoliubov inequality,
\be
F\le F_0+\langle H-H_0\rangle_0\,,
\label{a-1}
\ee
where $H$ is the exact DGM Hamiltonian and $H_0$ is that of a reference system (here an Einstein crystal with the same underlying lattice $\{{\bf R}_k\}$ as for the given DGM crystal). The average $\langle\ldots\rangle_0$ is taken over the canonical distribution pertaining to $H_0$. The Helmholtz free energy of the reference system reads:
\be
F_0=Nk_BT\left(\frac{3}{2}\ln\frac{\alpha}{\pi}+3\ln\frac{\Lambda}{\sigma}\right)\,,
\label{a-2}
\ee
where $\alpha=\beta c\sigma^2/2$, $c$ being the spring constant, and $\beta=(k_BT)^{-1}$. It is then possible to show (for $\sigma_2=\sigma$) that
\be
\langle H-H_0\rangle_0=\frac{1}{2}\epsilon N\sum_{k\neq 1}I_k-\frac{1}{2}\epsilon_2N\sum_{k\neq 1}J_k-\frac{3}{2}Nk_BT\,,
\label{a-3}
\ee
where
\be
I_k=\left(\frac{\alpha}{2+\alpha}\right)^{3/2}\exp\left\{-\frac{\alpha}{2+\alpha}\frac{({\bf R}_k-{\bf R}_1)^2}{\sigma^2}\right\}
\label{a-4}
\ee
while the calculation of $J_k$ is deferred to the end of the appendix. Within the VT, the optimal estimate of the crystal free energy for the given $T$ and $\rho$ is the minimum of the right-hand side of Eq.\,(\ref{a-1}) as a function of $\alpha$. As far as the fluid phase is concerned, its pressure $P$ and chemical potential $\mu$ are given by closed-form expressions in terms of the radial distribution function and of the direct correlation function in the HNC approximation, which is known to be accurate for all densities at not too low temperature~\cite{Likos}. The exact location of the transition points is found by looking for intersections between the various $\mu(P)$ branches at constant temperature or, equivalently, by the common-tangent construction. Finally, the liquid-vapor binodal line can roughly be located at the boundary of the $(\rho,T)$ region where the iterative procedure for solving the HNC equation fails to converge.

An important difference with respect to the GCM is the partitioning of the solid sector, which in the DGM comprises as many as four crystalline phases (i.e., the same as those quoted in Table I). Upon cooling, the VT predicts an abrupt transition (close below $T=0.0015\,\epsilon/k_B$) from a situation where the bcc crystal is metastable with respect to both the liquid and fcc phases to a bcc-fcc equilibrium with no stable liquid phase on the low-density side (see Fig.\,4). It goes without saying that, on the specific question of the relative stability between the liquid and the low-pressure bcc phase as well as in all cases where tiny free-energy differences are involved (e.g. the solid-solid equilibria), only numerical simulation supplemented with ``exact'' free-energy calculations can say a final word.

Looking retrospectively at the ``exact'' results (see Fig.\,9), the predictions of the variational theory (VT) are pretty good, even though (i) it overestimates $T_{\rm max}$ by about 20\% (the errors are even larger at higher pressures) and (ii) no hcp-liquid coexistence locus is actually present in the DGM.
As to the low-pressure region of the phase diagram, where according to our calculations a number of oddities occur for the DGM, none of these features are anticipated by the VT for the likely reason that the HNC approximation is poor at very low temperatures.

In closing, we shortly outline the calculation of the integral
\ba
J_k&=&\frac{\int{\rm d}^3r_1{\rm d}^3r_k\,\exp\left\{-(|{\bf r}_1-{\bf r}_k|-\xi)^2/\sigma^2\right\}\exp\left\{-\beta c\left[({\bf r}_1-{\bf R}_1)^2+({\bf r}_k-{\bf R}_k)^2\right]/2\right\}}{\int{\rm d}^3r{\rm d}^3r'\,\exp\left\{-\beta c(r^2+r'^2)/2\right\}}
\nonumber \\
&\equiv&\left(\frac{\beta c}{2\pi}\right)^3{\cal I}_k\,,
\label{a-5}
\ea
which occurs in the evaluation of the VT approximant (\ref{a-1}). Upon changing integration variables from ${\bf r}_1,{\bf r}_k$ to ${\bf r}\equiv{\bf r}_1-{\bf r}_k,{\bf r}'\equiv{\bf r}_1-{\bf R}_1$ one readily obtains:
\be
{\cal I}_k=\int{\rm d}^3r{\rm d}^3r'\,\exp\{-(r-\xi)^2/\sigma^2\}\exp\left\{-\beta c\left[r'^2+({\bf r}'-{\bf r}+{\bf R}_1-{\bf R}_k)^2\right]/2\right\}\,,
\label{a-6}
\ee
the Jacobian of the transformation being 1. Denoting ${\bf Q}_k={\bf r}-{\bf R}_1+{\bf R}_k$, ${\cal I}_k$ can further be simplified to
\be
{\cal I}_k=\int{\rm d}^3r\,\exp\left\{-\frac{(r-\xi)^2}{\sigma^2}-\frac{1}{2}\beta cQ_k^2\right\}\int{\rm d}^3r'\,\exp\left\{-\beta c(r'^2-{\bf Q}_k\cdot{\bf r}')\right\}\,.
\label{a-7}
\ee
The inner integral in (\ref{a-7}) is the classical Gaussian integral; its value is $(\beta c/\pi)^{-3/2}\exp\{\beta cQ_k^2/4\}$. Hence, we find:
\be
{\cal I}_k=\left(\frac{\pi}{\beta c}\right)^{3/2}\int{\rm d}^3r\,\exp\left\{-\frac{(r-\xi)^2}{\sigma^2}\right\}\exp\left\{-\frac{1}{4}\beta c({\bf r}-{\bf R}_1+{\bf R}_k)^2\right\}\,.
\label{a-8}
\ee
This integral is best computed in spherical coordinates, choosing the $z$ axis in the direction of ${\bf R}_1-{\bf R}_k$. The calculation is straightforward; using
\be
\int_0^\infty{\rm d}r\,re^{-ar^2+br}=\frac{1}{2a}+\frac{b}{4a}\sqrt{\frac{\pi}{a}}\,{\rm erfc}\left(-\frac{b}{2\sqrt{a}}\right)\exp\left\{\frac{b^2}{4a}\right\}\,,
\label{a-9}
\ee
the end result is (cf. Eq.\,(\ref{a-5}))
\ba
J_k&=&\frac{1}{8\pi}\sqrt{\frac{\beta c}{\pi}}\left(\frac{4\pi}{\beta c+4/\sigma^2}\right)^{3/2}\frac{\exp\left\{-\beta c({\bf R}_1-{\bf R}_k)^2/4-\xi^2/\sigma^2\right\}}{|{\bf R}_1-{\bf R}_k|}
\nonumber \\
&\times&\left[\left(\frac{2\xi}{\sigma^2}+\frac{\beta c}{2}|{\bf R}_1-{\bf R}_k|\right){\rm erfc}\left(-\frac{2\xi/\sigma^2+\beta c|{\bf R}_1-{\bf R}_k|/2}{\sqrt{\beta c+4/\sigma^2}}\right)\exp\left\{\frac{(2\xi/\sigma^2+\beta c|{\bf R}_1-{\bf R}_k|/2)^2}{\beta c+4/\sigma^2}\right\}\right.
\nonumber \\
&-&\left.\left(\frac{2\xi}{\sigma^2}-\frac{\beta c}{2}|{\bf R}_1-{\bf R}_k|\right){\rm erfc}\left(-\frac{2\xi/\sigma^2-\beta c|{\bf R}_1-{\bf R}_k|/2}{\sqrt{\beta c+4/\sigma^2}}\right)\exp\left\{\frac{(2\xi/\sigma^2-\beta c|{\bf R}_1-{\bf R}_k|/2)^2}{\beta c+4/\sigma^2}\right\}\right]\,,
\nonumber \\
\label{a-10}
\ea
which can also be presented as
\ba
J_k&=&\frac{1}{2\alpha}\left(\frac{\alpha}{2+\alpha}\right)^{3/2}\exp\left\{-\frac{\alpha}{2}\left(\frac{X_k}{\sigma}\right)^2-\left(\frac{\xi}{\sigma}\right)^2\right\}
\nonumber \\
&\times&\left[\left(\frac{2\xi}{X_k}+\alpha\right){\rm erfc}\left(-\frac{2\xi/\sigma+\alpha X_k/\sigma}{\sqrt{2(2+\alpha)}}\right)\exp\left\{\frac{(2\xi/\sigma+\alpha X_k/\sigma)^2}{2(2+\alpha)}\right\}\right.
\nonumber \\
&-&\left.\left(\frac{2\xi}{X_k}-\alpha\right){\rm erfc}\left(-\frac{2\xi/\sigma-\alpha X_k/\sigma}{\sqrt{2(2+\alpha)}}\right)\exp\left\{\frac{(2\xi/\sigma-\alpha X_k/\sigma)^2}{2(2+\alpha)}\right\}\right]
\label{a-11}
\ea
with $X_k=|{\bf R}_k-{\bf R}_1|$. It is easy to check that $J_k$ reduces to $I_k$ when $\xi$ vanishes, as it should do.

\newpage
\section*{Figure captions}
%
%
{\bf FIGURE 1}. Top: DGM potential $u(r)$, see Eq.\,(\ref{2-1}), for $\epsilon_2=0.005\epsilon,\sigma_2=\sigma$, and $\xi=3\sigma$ (solid line), together with the force $f(r)=-u'(r)$ (dashed line). In the lower panel, a magnification of the small-$u$ region is shown so as to highlight the attractive well.

%
%
{\bf FIGURE 2}. (Color online). Numerical DGM phase diagram in the $P$-$T$ plane. We show: the liquid-vapor coexistence points obtained from Gibbs-ensemble simulations (black diamonds), along with the estimated critical point (black asterisk); the solid-liquid and solid-solid coexistence points obtained from the ``exact'' free-energy calculations described in the text (filled dots, squares, and triangles); a few points on the structural-anomaly locus (red open diamonds) and one point on the volumetric-anomaly locus (the red open square; the maximum-density locus continues towards higher densities in a region of the phase diagram that we have not investigated). The lines through the points are plotted as a guide for the eye. The bcc-vapor coexistence line is only schematic. The black open triangle lies on the metastable continuation of the high-pressure fcc-bcc locus inside the liquid region of stability. We included it only to show that the fcc-bcc coexistence temperature reaches a maximum as a function of pressure roughly at the confluence with the melting line.

%
%
{\bf FIGURE 3}. (Color online). Numerical DGM phase diagram in the $\rho$-$T$ plane. We show: a number of points along the liquid-vapor binodal line obtained by Gibbs-ensemble simulations (black diamonds), along with the estimated critical point (black asterisk); a number of points along the line enclosing the region where the HNC equation could not be solved (blue diamonds joined by a dotted line); the solid-liquid and solid-solid coexistence points obtained from the ``exact'' free-energy calculations described in the text (filled dots, squares, and triangles); some points on the $P=0$ isobar of the bcc solid (black triangles pointing to the right joined by a dotted line). All the lines through the points are plotted as a guide for the eye. Note that the phase-diagram region enclosed in the red frame is shown magnified in the next Fig.\,4.

%
%
{\bf FIGURE 4}. (Color online). DGM phase diagram in the $\rho$-$T$ plane: Magnification of the low-density region. We show: one point on the liquid-vapor binodal line obtained by Gibbs-ensemble simulations (black diamond); a number of solid-liquid and solid-solid coexistence points obtained from the ``exact'' free-energy calculations described in the text (filled dots, squares, and triangles); some points on the bcc sublimation line, obtained from MC simulations of the bcc crystal at $P=0$ (black triangles pointing to the right joined by a dotted line). All the lines through the points are plotted as a guide for the eye. The yellow shaded regions denote two-phase coexistence regions.

%
%
{\bf FIGURE 5}. (Color online). Radial distribution function (RDF) for a number of temperatures ($T=0.003,0.004,\ldots,0.010$) along the $P=0.002$ isobar: GCM fluid (red curves on the right) and the DGM system under study (black curves on the left). The arrow marks the direction of temperature increase.

%
%
{\bf FIGURE 6}. (Color online). Schematic phase diagrams (see text) for the DGM system as a function of $\epsilon_2$, for fixed $\sigma_2=\sigma$ and $\xi=3\sigma$ (same symbols and notation as in Fig.\,8). Observe that the actual melting temperatures are about a half of those shown.

%
%
{\bf FIGURE 7}. (Color online). Schematic phase diagrams (see text) for the DGM system as a function of $\xi$, for fixed $\epsilon_2=0.005\epsilon$ and $\sigma_2=\sigma$ (same symbols and notation as in Fig.\,8). Observe that the actual melting temperatures are about a half of those shown.

%
%
{\bf FIGURE 8}. (Color online). Schematic phase diagrams of the GCM fluid (left) and of the DGM system under study (right). These diagrams were obtained by using the HNC approximation in combination with a Lindemann-type criterion of crystal melting\,\cite{Prestipino3} (see text). The $(\rho,T)$ set of points where the HNC equation has no solution is the region delimited by the black dots. The lines are tentative melting loci for the fcc (blue dotted line), the bcc (red solid line), and the hcp crystal (cyan dashed line); more realistically, the actual values of $T_m$ are roughly a half of those shown. A narrow bcc basin is expected to occur for the DGM system at low density, bounded from above by a reentrant-melting line. The more standard GCM-type reentrant melting is located at much higher densities. Another fcc basin at lower densities would never enter into play since being hidden under the liquid-vapor spinodal region. In the present DGM system, the hcp crystal would apparently be stable only at the lowest temperatures.

%
%
{\bf FIGURE 9}. (Color online). Numerical DGM phase diagram in the $P$-$T$ plane: comparison between the ``exact'' coexistence loci of Fig.\,2 (black solid lines) and the approximate results of the theoretical approaches described in the appendix. We show: the HUIM melting points obtained by heating (in steps of $\Delta T=0.0001$) the solid system along isobaric paths until it melted (blue crosses joined by a dashed line); the VT coexistence points (blue open dots and triangles joined by dotted lines). The lines through the points are plotted as a guide for the eye.

%
%
\begin{figure}
\centering
\includegraphics[width=16cm]{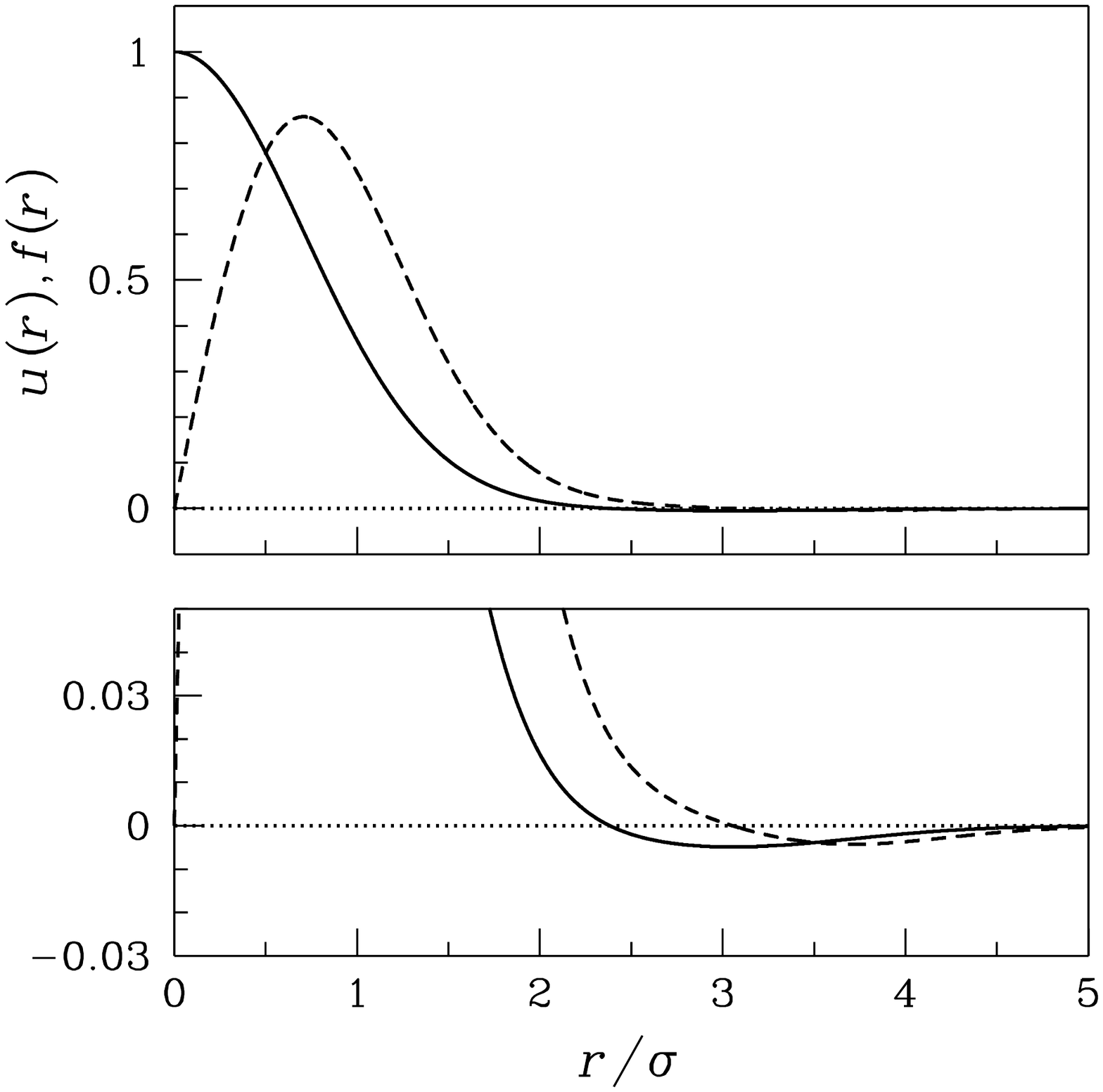}
\caption{}
\label{fig1}
\end{figure}

%
%
\begin{figure}
\centering
\includegraphics[width=16cm]{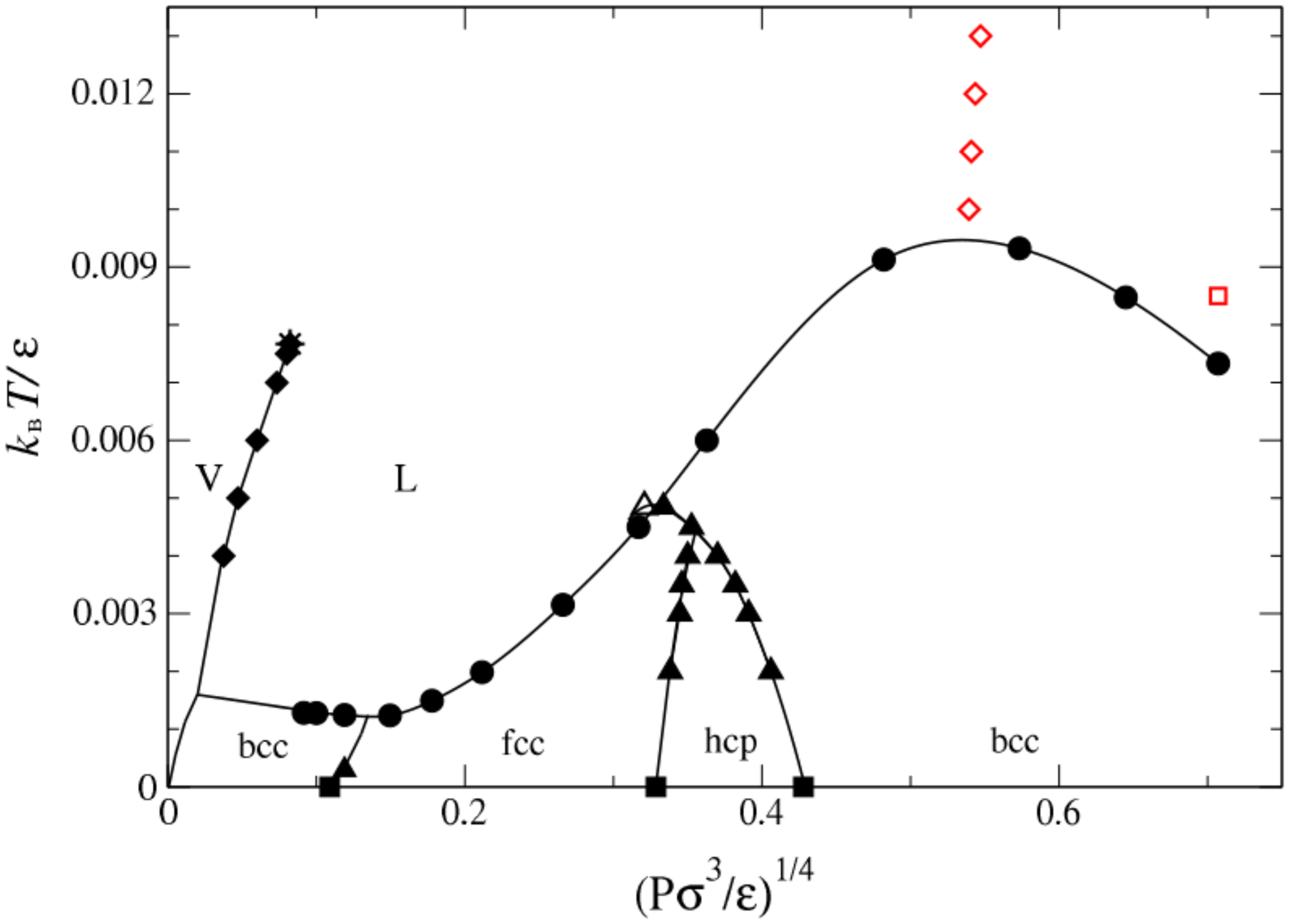}
\caption{}
\label{fig2}
\end{figure}

%
%
\begin{figure}
\centering
\includegraphics[width=16cm]{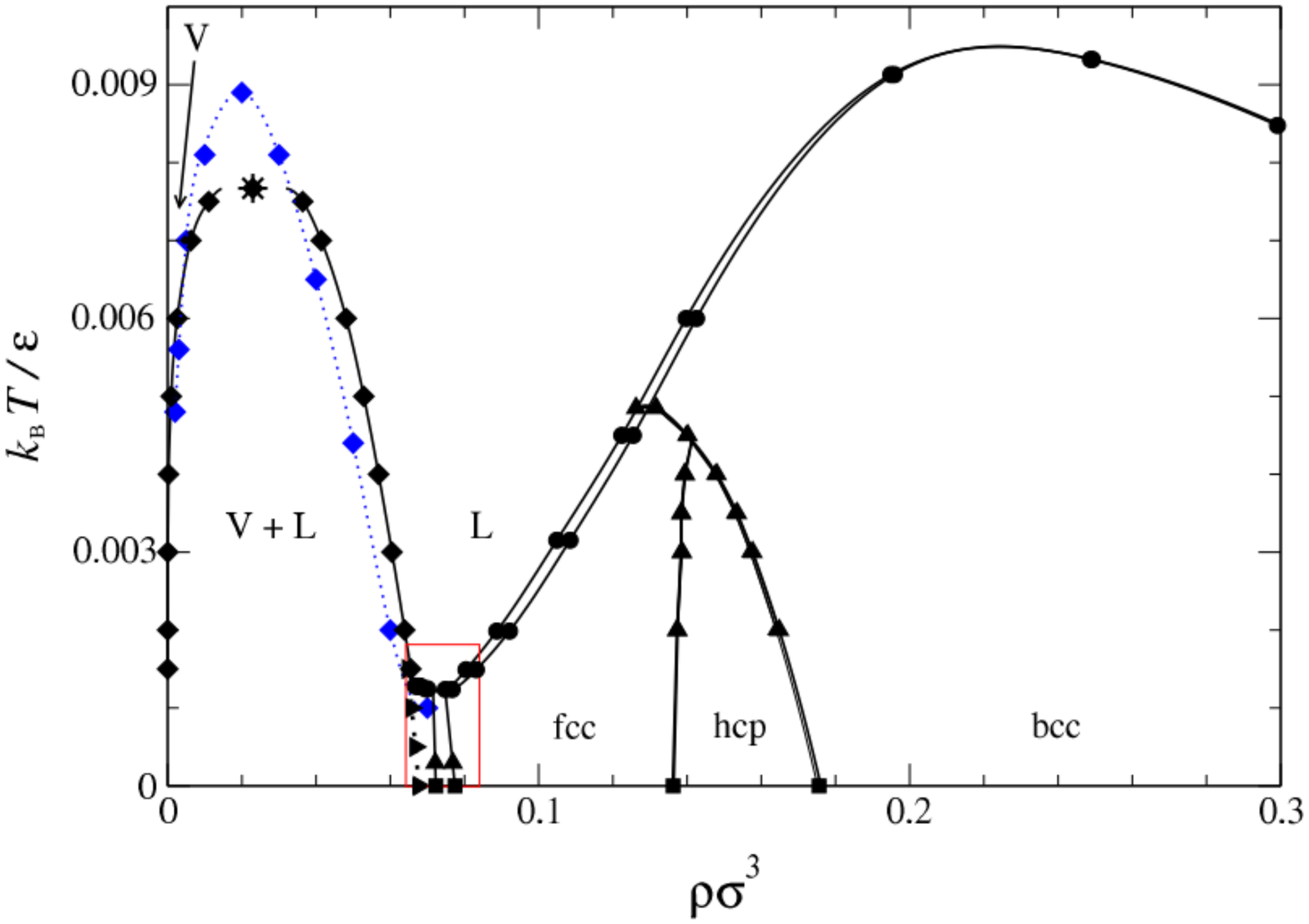}
\caption{}
\label{fig3}
\end{figure}

%
%
\begin{figure}
\centering
\includegraphics[width=16cm]{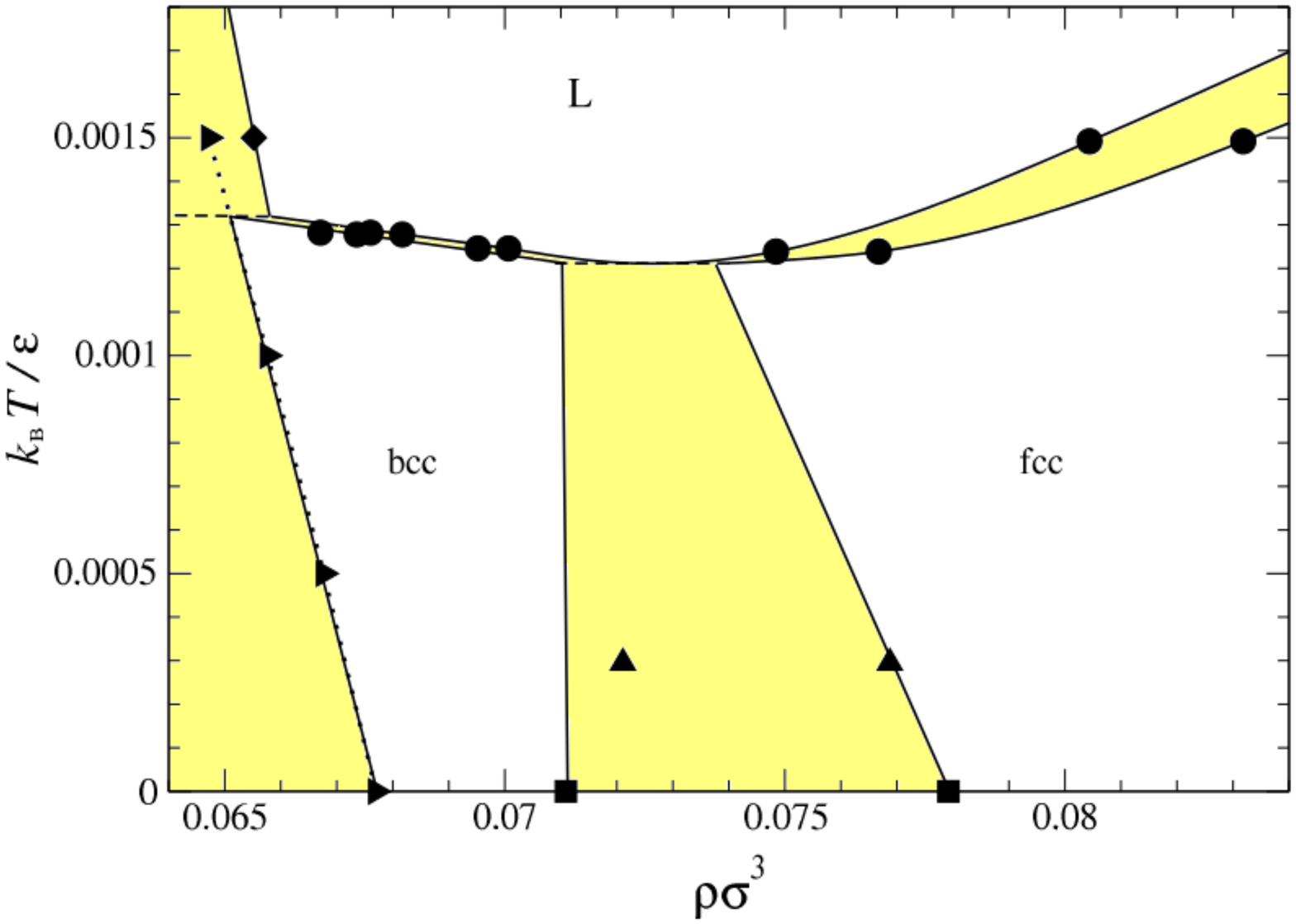}
\caption{}
\label{fig4}
\end{figure}

%
%
\begin{figure}
\centering
\includegraphics[width=16cm]{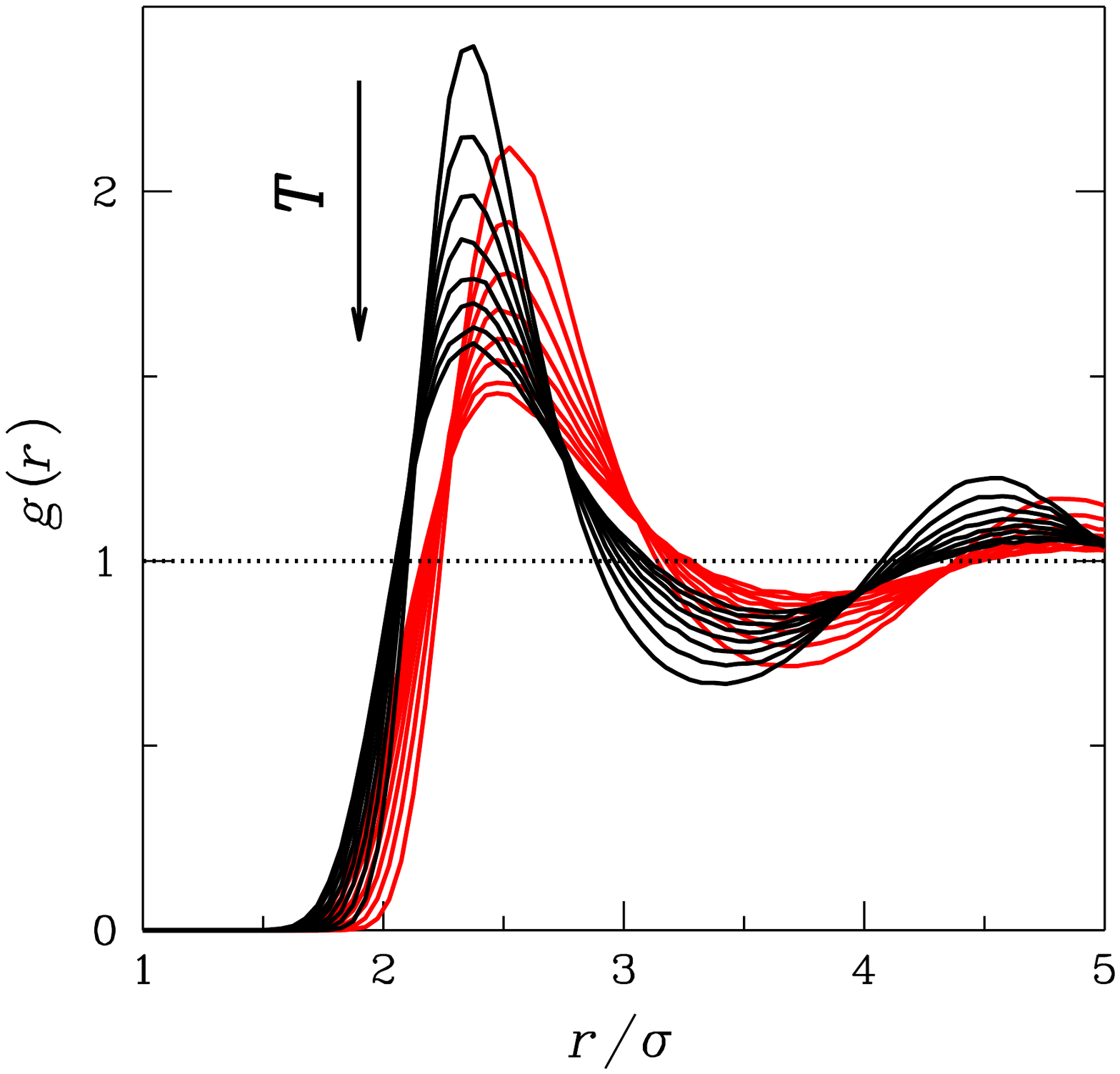}
\caption{}
\label{fig5}
\end{figure}

%
%
\begin{figure}
\centering
\includegraphics[width=16cm]{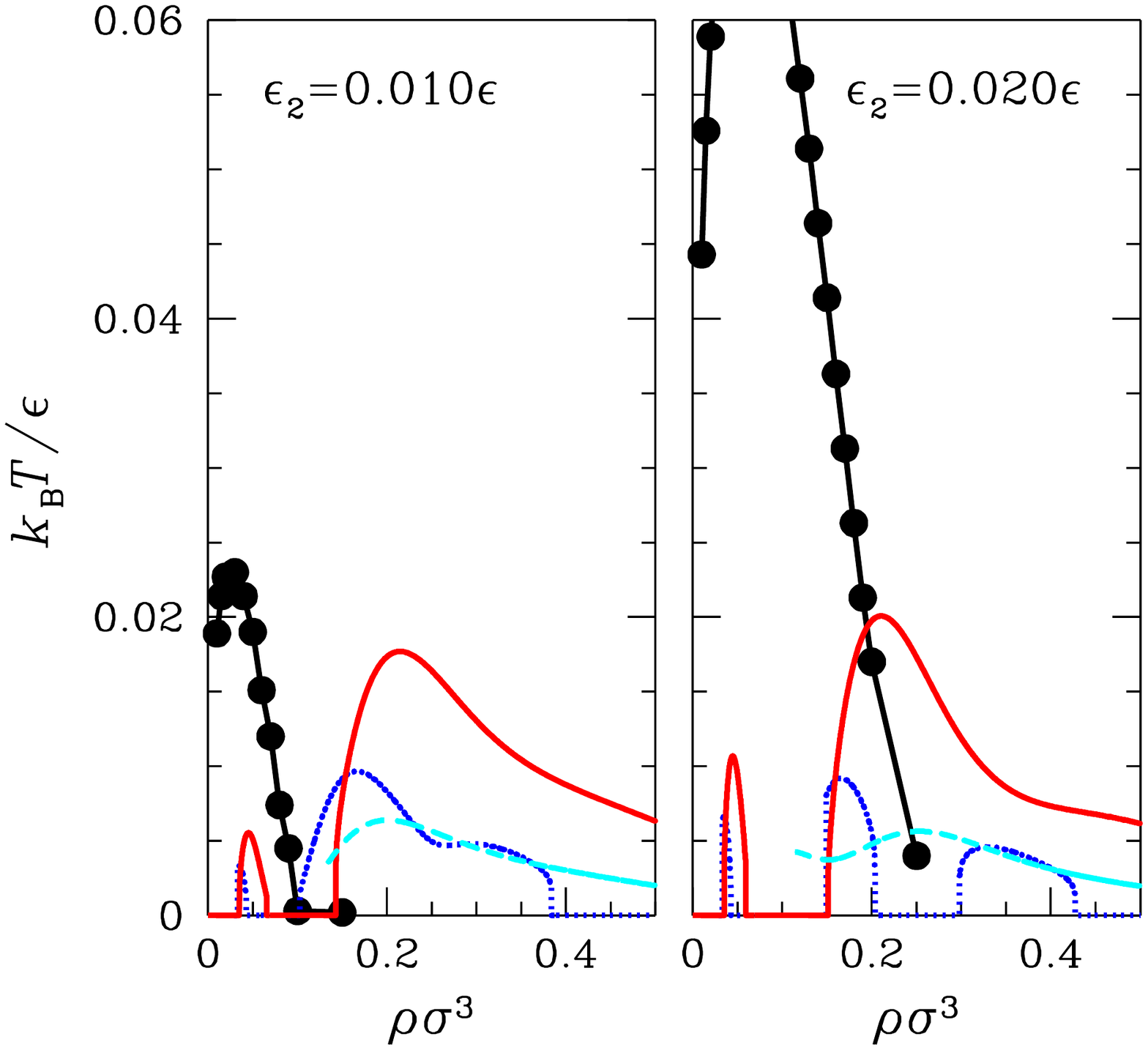}
\caption{}
\label{fig6}
\end{figure}

%
%
\begin{figure}
\centering
\includegraphics[width=16cm]{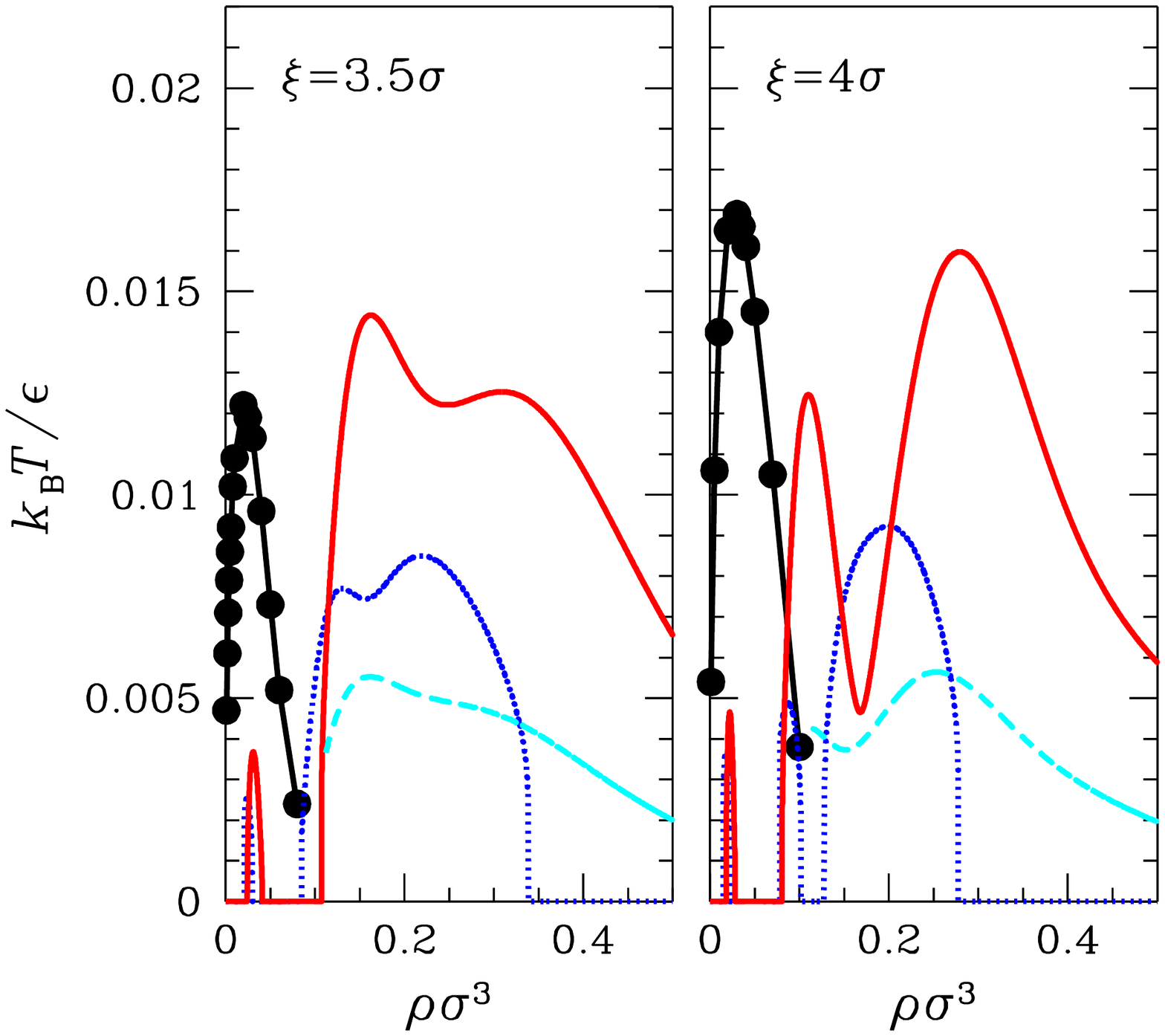}
\caption{}
\label{fig7}
\end{figure}

%
%
\begin{figure}
\centering
\includegraphics[width=16cm]{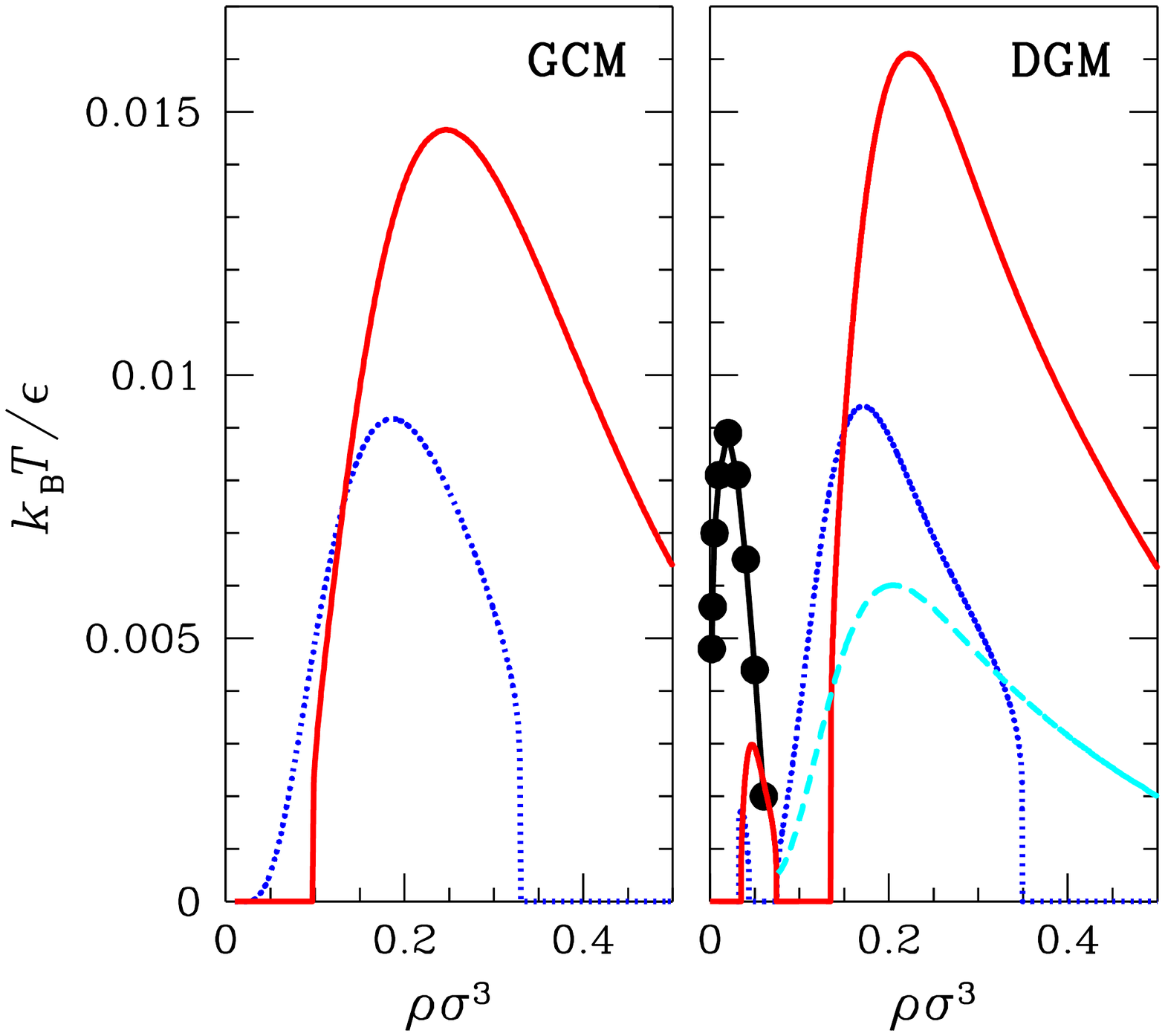}
\caption{}
\label{fig8}
\end{figure}

%
%
\begin{figure}
\centering
\includegraphics[width=16cm]{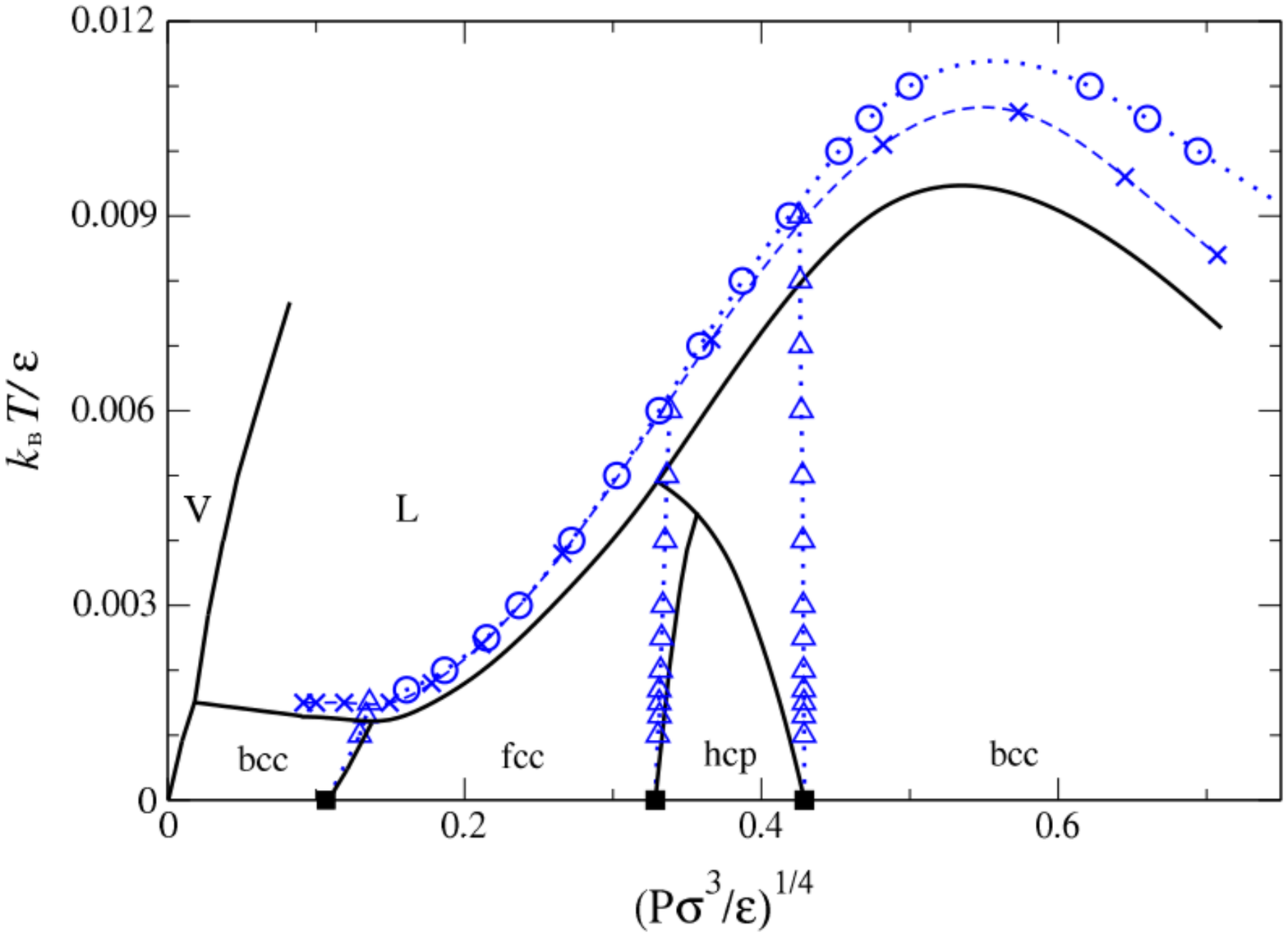}
\caption{}
\label{fig9}
\end{figure}
\end{document}